\documentclass[onecolumn,12pt]{article}
\usepackage[totalwidth=480pt,totalheight=640pt]{geometry}
\usepackage[centertags]{amsmath}
\usepackage{array,multirow}
\usepackage{graphicx}
\usepackage{color}
\usepackage{setspace}
\usepackage{amsmath,slashed}
\usepackage{amssymb}

\usepackage{microtype}
\pdfoutput=1

\def\ltap{\ \raise.3ex\hbox{$<$\kern-.75em\lower1ex\hbox{$\sim$}}\ }
\def\gtap{\ \raise.3ex\hbox{$>$\kern-.75em\lower1ex\hbox{$\sim$}}\ }
\def\fbpartial{\ \raise1.2ex\hbox{\kern-0.3em$\leftrightarrow$\kern-.85em\lower1.5ex\hbox{$\partial$}\kern-.25em}\ }

\newcommand{\gsim}{\lower.7ex\hbox{$\;\stackrel{\textstyle>}{\sim}\;$}}
\newcommand{\lsim}{\lower.7ex\hbox{$\;\stackrel{\textstyle<}{\sim}\;$}}
\def\OO{{\cal O}}
\def\CC{{\cal C}}

\def\LL{{\cal L}}

\def\unit{\relax{\rm 1\kern-.26em I}}
\newcommand{\half}{{\frac{1}{2}  }}

\newcommand{\hc}{\text{ h.c. }}

\newcommand{\Tr}{{\text{ Tr }}}

\newcommand{\be}{\begin{eqnarray}}
\newcommand{\ee}{\end{eqnarray}}
\newcommand{\nn}{\nonumber}

\newcommand{\tha}{\vartheta_{\text{a}}}
\newcommand{\delfb}{\overset{\leftrightarrow}{\nabla}}

\newcommand{\INT}{ \int\!\! \frac{d \tilde \omega}{2\pi} \frac{d^2  \tilde k}{(2\pi)^2}\;\;}

\newcommand\lrnab{\raise .8ex\hbox{$^\leftrightarrow$} \hspace{-8.8pt}
\nabla}
\newcommand\lnab{\raise .8ex\hbox{$^\leftarrow$} \hspace{-8.8pt}
\nabla}
\newcommand\rnab{\raise .8ex\hbox{$^\rightarrow$} \hspace{-8.8pt}
\nabla}

\usepackage{hyperref}

\begin{document}

\begin{titlepage}

\begin{center}
\vspace{1cm}

{\Large \bf  A New Theory of Anyons }


\vspace{0.8cm}

\small
\bf{A. Liam Fitzpatrick$^1$, Shamit Kachru$^{1,2}$, Jared Kaplan$^2$, Emanuel Katz$^{1,2,3}$, Jay G. Wacker$^2$}
\normalsize

\vspace{.5cm}

{\it $^1$ Stanford Institute for Theoretical Physics, Stanford University, Stanford, CA 94305}\\
{\it $^2$ SLAC National Accelerator Laboratory, 2575 Sand Hill, Menlo Park, CA 94025} \\
{\it $^3$ Physics Department, Boston University, Boston, MA 02215} \\

\end{center}

\vspace{1cm}

\begin{abstract}

We study a 2+1 dimensional theory of bosons and fermions with an $\omega \propto k^2$ dispersion relation.  The most general interactions consistent with specific symmetries impart fractional statistics to the fermions. Unlike examples involving Chern-Simons gauge theories, our statistical phases derive from the exchange of gapless propagating bosons with marginal interactions.   Even though no gap exists, we show that  the anyonic statistics are precisely defined.  Symmetries combine with the vacuum structure to guarantee the non-renormalization of our anyonic phases.

\end{abstract}

\bigskip

\end{titlepage}

\section{Introduction}

In two spatial dimensions particles can be `anyons', with quantum statistics that interpolate between Bose-Einstein and Fermi-Dirac \cite{leinaas,wilczek, Rao:1992aj}.  Anyonic statistics can be Abelian or non-Abelian and arise when multiparticle wavefunctions fall into non-trivial representations of the braid group on their spacetime trajectories.   Despite significant interest in their behavior, calculable field theories of anyons have been limited to a single class where the anyonic phases derive from interactions with a Chern-Simons gauge field \cite{Zee}.  Given the striking properties of anyons and the essential role that statistics plays in quantum many-body systems, it is worth  asking whether Chern-Simons anyons are generic, or whether qualitatively different field theories of anyons may be found.  

This article will present a new quantum field theory of anyons which has only marginal interactions.  
The theory consists of a non-relativistic fermion $\psi$ and a real scalar field $\phi$ with dispersion relation $\omega \propto k^2$, often referred to as `$z=2$'.  
A novel symmetry structure enforces these properties.  The most important  symmetry transforms the fields as
\begin{eqnarray}
\phi(t, \vec{x}) \to \phi(t, \vec{x}) - v( \vec{x})\qquad \psi(t,\vec{x}) \to \exp(i u(\vec{x}))\,\psi(t,\vec{x})\qquad\text{with} \quad \nabla_i u = \epsilon_{ij} \nabla^j v .
\end{eqnarray} 
The pair of functions $(u, v)$ satisfy the Cauchy-Riemann equations, so we will refer to these transformations as the Cauchy-Riemann symmetry.
This symmetry guarantees a $z=2$ kinetic term and derivative couplings for $\phi$.  

The anyonic statistics for $\psi$ arise from the interactions
\be
  -g \nabla^2 \phi \psi^\dagger \psi  \quad \text{ and } \quad -i \alpha  \vec{\nabla}\phi \times \psi^\dagger \delfb\psi .
  \label{eq:interactions} 
 \ee
 The reader may find it useful to interpret these couplings in terms of the ``dual'' gauge field $a^i = \epsilon^{ij} \nabla_j \phi$, in which case the first is a coupling of $\psi^\dag \psi$ to the ``magnetic'' field $b=\epsilon^{ij} \nabla_i a_j$ and the latter is the standard coupling  $a^i J_i$.  As we will see, the analogy with $a_i$ is no accident; the Cauchy-Riemann symmetry has forced $\epsilon_{ij} \nabla^j \phi$ to couple in a gauge-invariant way.  As one $\psi$ particle completes an orbit around another, the coupling to $\phi$ generates an anyonic phase of $e^{i \tha}$ with 
\be
\tha =  g \alpha .
\ee
Remarkably, this anyonic phase and the individual couplings $\alpha$ and $g$ are not renormalized at any order in perturbation theory.  This follows as a highly non-trivial consequence of the Cauchy-Riemann symmetry combined with the absence of $\psi$ anti-particles in the vacuum.
A non-Abelian generalization of the classical theory can also be constructed, where $\phi$ and $\psi$ transform in representations of a non-Abelian group.

Our theory differs from Chern-Simons theory in several important respects.  One major difference is that the $\phi$ theory is not topological, and in particular, $\phi$ is a gapless propagating degree of freedom\footnote{Despite the absence of a gap, the anyonic phase is well-defined, as we show in section \ref{Sec: DefinitionOfStatistics}}.  As a consequence,  the motion of $\psi$ particles can produce $\phi$ radiation at arbitrarily low energies.  Although $\phi$ radiation is a significant effect, remarkably, there are no IR divergences associated with the amplitudes for emission of soft $\phi$ particles.  This sharply distinguishes our theory from models of Yukawa-type interactions between gapless scalars and fermions at a Fermi surface in 2+1 dimensions, which suffer from  well-known IR divergences that obscure their low-energy behavior  \cite{SungSik}.

In the atomic and condensed matter physics literature, $\psi$-like particles appear in theories of trapped cold atoms \cite{AMO}, and close relatives (with kinematics that have very important differences) appear in theories of `quadratic band touching' \cite{QBT}.  Theories of $z=2$ scalars like our $\phi$ field have been discussed in \cite{Lifscalar}.  However, we will be studying various properties of our $\phi, \psi$ system largely without reference to any specific implementation in a realistic substance.

The outline of the paper is as follows.  In  \S\ref{sec:SemiClassicalAnyons} we explain how the theory works at a semi-classical level, including the symmetries and the basics of the anyonic phase.  We also explain the relationship of our theory to Chern-Simons theory, including a remarkable `Mulligan Duality' with the critical Lifshitz-Chern-Simons theory \cite{Mulligan:2010wj}.  Then in  \S\ref{sec:QuantumAnyons} we analyze the theory at the full quantum level.  We explain why the symmetries and dynamics of the theory lead to the non-renormalization of the couplings $\alpha$ and $g$, and then we check these statements with explicit one-loop computations.  We also show that there are no IR divergences in our theory.  In section \ref{sec:NonAbelian} we sketch how the theory can be generalized to produce non-Abelian anyons.  Finally in  \S\ref{sec:Discussion} we briefly discuss the prospects for finding experimental realizations of our theory and future directions. 

\section{A Semi-Classical Theory of Anyons}
\label{sec:SemiClassicalAnyons}

In this section we will display the symmetries of our model and explain how it gives rise to anyonic statistics at the semi-classical level. 
We will begin by studying a $2+1$ dimensional theory coupling a boson $\phi$ to a one-component 
fermion $\psi$ at zero chemical potential with the action  
\be
S_A=\int dt d^2x \left[ \frac{1}{2} \left( \dot \phi^2 -  (\nabla^2 \phi)^2\right)  -  \psi^\dag  i\partial_t  \psi - \gamma \left| \left(\nabla_i + i\alpha \epsilon_{ij} \nabla^j \phi \right) \psi \right|^2 + g \nabla^2 \phi  \psi^\dag \psi \right].
\label{eq:ourLag}
\ee
The action has been chosen so that both fields have a dispersion relation $\omega \propto \vec k^2$, often referred to as $z=2$.  Below, we will show that the interactions cause $\psi$ to develop an anyonic phase.  As we scale energy $\omega \rightarrow s \omega$, all of the parameters appearing in the action are marginal if we additionally choose the scalings
\be
dt \rightarrow s^{-1}dt , \ \ \ dx \rightarrow s^{-\frac{1}{2}} dx, \ \ \ \phi \rightarrow s^0 \phi, \ \ \ \psi \rightarrow s^{\frac{1}{2} } \psi,
\ee
in accord with the kinetic terms for $\phi$ and $\psi$.  The $\phi$ field is dimensionless, much like a boson field in $1+1$ dimensions with a relativistic dispersion relation.  Naively this could give rise to an infinite number of relevant and marginal couplings, but these  will be forbidden by shift symmetries on $\phi$. 
The parameter $\gamma$ sets the relative `mass' of the non-relativistic  $\psi$ particles relative to the $\phi$ particles.  A quartic coupling in $\psi$ would vanish due to the non-bosonic statistics.

Other than an allowed chemical potential term $\mu \psi^\dagger \psi$ that has been tuned to zero, this is the most general renormalizable action consistent
with the following somewhat unorthodox set of symmetries.  Consider first a holomorphic function
\be
f(z) = u(x,y) + i v(x,y),
\label{eq:f}
\ee
where $z = x+iy$.   Then as usual, $u$ and $v$ will be harmonic functions related by the Cauchy-Riemann equations  
\be
\nabla_i u = \epsilon_{ij} \nabla^j v .
\ee
which follow from $\partial_{\bar z} f=0$.  For any such $u$ and $v$, our Lagrangian will be invariant under the gauge transformation
\be
\psi \to e^{i \alpha u} \psi \ \ \ \mathrm{and} \ \ \ \phi \to \phi -  v .
\ee
For spatially varying $u$ and $v$ we interpret these symmetries as a redundancy of the description; but transformations with constant $u$ and $v$ are global symmetries of our theory. A quick calculation with Noether's theorem shows that the global symmetry associated with constant $u$ generates the following number current:
\be
J_N^0 = \psi^\dagger \psi , && J_N^i=\gamma  \left( - i \psi^\dagger \lrnab^i \psi +2\alpha \epsilon^{ij} \nabla^j\phi (\psi^\dagger \psi) \right) =-i\gamma \psi^\dag \overset{\leftrightarrow}{\mathcal{D}^i} \psi  ,
\label{eq:numbercurrent}
\ee
written in terms of the covariant derivative $\mathcal{D}^i = \nabla_i + i\alpha \epsilon_{ij} \nabla_j \phi$, whereas the symmetry associated with constant $v$ generates 
\be
J_\phi^0 = \dot{\phi}  , && J_\phi^i = \left( - \nabla^i \nabla^2 \phi -\alpha\epsilon^{ij} J_N^j + g \nabla^i (\psi^\dagger \psi) \right) .
\label{eq:phicurrent}
\ee

 This is not yet enough to forbid a possible $\dot{\phi} \psi^\dagger \psi$ coupling, so let us impose on our theory an invariance under an additional global symmetry, where $\phi$ shifts as
\be
\phi &\rightarrow& \phi + c t ,
\ee
with $c$ constant.  Under this transformation the shift of the $\dot \phi^2$ kinetic term is a total derivative, but the $\dot \phi \psi^\dag \psi$ interaction is forbidden.  All other interactions are manifestly invariant under this additional symmetry.  

The theory produces anyonic statistics for $\psi$ through an interplay between the $g$ and $\alpha$ coupings.  For this purpose we will study the behavior of a state with two $\psi$ particles as we slowly move one particle around the other.  Each $\psi$ particle will source the $\phi$ field, which has the equation of motion
\be
\ddot \phi + \nabla^4 \phi = g \nabla^2 (\psi^\dag \psi) - \alpha \epsilon_{ij} \nabla^i J_N^j.
\ee
  If we consider a state with a static $\psi$ density $\rho=\psi^\dagger\psi$, the Euler-Lagrange equations for $\phi$ give
\be
\nabla^4 \phi(\vec{x}) = g \nabla^2\psi^\dagger\psi(x) \quad \Rightarrow\quad
\nabla^2 \phi(x) =  g \rho(x) + f(x),
\ee
where $\nabla^2 f=0$. 
The boundary conditions for $\phi$ at infinity  require $f \to 0$ to avoid a large boundary contribution to the action.  
Thus  $\rho$ sources $\phi$ via the standard 2-d Laplace equation.  In particular, a $\psi$ particle at $\vec{x}=0$ gives rise to a long-ranged logarithmic $\phi$ field
\begin{eqnarray}
\phi(\vec{x}) =  \frac{g}{2 \pi} \log |\vec{x}| .
\end{eqnarray}

When a spectator $\psi$ particle moves through this background $\phi$ field, it accumulates a phase.  The easiest way to determine this phase is to study the first-quantized action for $\psi$ particles.  In App.~\ref{FirstQuantized} we derive the one-particle action
\be
S_\psi =  \int d \tau \left(\frac{\dot x^2(\tau)}{4 \gamma}  + g \nabla^2 \phi(x(\tau)) - \alpha  \dot x_i(\tau) \epsilon^{ij} \nabla_j \phi(x(\tau)) \right).
\ee
Now imagine moving one $\psi$ particle about another in a counter-clockwise circle.  Using Stokes' theorem and the equations of motion for the $\phi$ field, we find an accumulated phase
\be
\tha=-i \alpha \oint_{\partial M}  d \theta \left( \hat \theta_i \epsilon^{ij} \nabla_j \phi \right)  &=& i \alpha \int_{M} d^2 x (\nabla^2 \phi ) = i g \alpha \int_{M} d^2 x \ \!  \rho(x) .
\ee
So if the trajectory of a $\psi$ particle encloses one other $\psi$ particle, we pick up an anyonic phase $i g \alpha$.   At the semi-classical level this phase depends only on the charge enclosed by the path, and hence on its homotopy class as a path in the punctured plane with locations of $\psi$-particles removed.  We will see in the next section that this continues to hold when quantum corrections are included.

\subsection{Comparison with Chern-Simons and Lifshitz-Chern-Simons}
\label{sec:ComparisonChernSimons}

In our theory anyonic statistics arise from the exchange of a propagating $\phi$ field.  This contrasts sharply with the other known controlled effective theory of anyons, namely Chern-Simons theory \cite{Zee}.  For instance, $z=2$ anyons can be described with Chern-Simons theory by using the action
\be
S_{\text{CS}} = \int dt d^2 x     \left[ -\psi^\dag i D_t \psi - \gamma | D_i \psi |^2 +  \kappa \epsilon^{\mu \nu \rho} A_\mu \partial_\nu A_\rho \right],
\label{eq:CSaction}
\ee
where $D_\mu = \partial_\mu + i A_\mu$ is the usual gauge covariant derivative.  These $\psi$ particles source the topological Chern-Simons field via the $A_0$ equation of motion, which gives
\be
 B = \frac{\gamma}{\kappa} \psi^\dag \psi,
\ee
where the magnetic field $B = \epsilon^{ij} \nabla_i A_j$  as usual.  The $\psi$ particles pick up an anyonic phase when they encircle charge; this can be seen by studying their first quantized action, which includes the usual operator $\dot x_i A^i$.

Our theory is however related, modulo possible global and boundary effects, to a critical point of the Abelian $z=2$ Lifshitz-Chern-Simons theory \cite{Mulligan:2010wj} coupled to $\psi$:\footnote{We are grateful to Mike Mulligan for explaining this to us. }  
\be
S_{\textrm{Lif-CS}} &=& \int dt d^2 x \left[ -\psi^\dag i D_t \psi - \Gamma | D_i \psi |^2 +  \kappa \epsilon^{\mu \nu \rho} A_\mu \partial_\nu A_\rho 
\right. \nn\\ && \left.
  + E^i (\dot{A}_i - \nabla_i A_0) - c^2 k_0^2 E^2 - \frac{c^2}{2} (\nabla_i E_j)^2 - \frac{f^2}{2} (\nabla \times A)^2 \right].
 \label{eq:LifCS}
\ee
The $E^2$ term is the unique relevant interaction, and so for any non-vanishing value of $c^2k^2_0$, it dominates the low-energy behavior of the theory. 
In particular, when $c^2k_0^2>0$, one can integrate $E^i$ out and in the IR one just obtains the original Chern-Simons action  (\ref{eq:CSaction}) plus a Maxwell term, which gives
the $\psi$ fields an anyonic phase $\tha\sim \frac{1}{\kappa}$ as usual.    However, at the critical point $k^2_0=0$, there exists a gapless propagating mode, and the IR description differs drastically.  

To see how the IR at $k_0^2=0$ is given locally by our theory, let us decompose $A_i$ into a longitudinal piece $\chi$ and a transverse piece $\Phi$:
\be 
A_i = \epsilon_{ij} \nabla_j \Phi + \nabla_i \chi.
\ee
Then, the equations of motion for $A_0$ and $E_i$ completely fix them in terms of the other fields:
\be
\label{cov}
 E^i &=& 2 \kappa \nabla^i \Phi - \frac{1}{c^2} \nabla^{-2} \epsilon^{ij} \nabla^j \dot\Phi  - \nabla^{-2} \nabla^i \psi^\dagger \psi , \\
A_0 &=& \dot{\chi} - c^2 \psi^\dagger \psi + 2 c^2 \kappa \nabla^2 \Phi. 
\ee
Remarkably, substituting this back into the action  (\ref{eq:LifCS}), one obtains an action that is completely local in terms of the $\Phi,\chi$, and $\psi$ fields!  
In fact,  after rescaling $x^i$ and $\psi$, and making the identifications
\be
\alpha = \frac{c^{\frac{1}{2} }}{(f^2+ 4 c^2 \kappa^2)^{\frac{1}{4} }}, \qquad 
g = 
 \frac{\alpha}{\sqrt{1+\left(\frac{f}{2 c \kappa}\right)^2}}, \qquad
\gamma = \frac{\Gamma}{c \sqrt{f^2 + 4 c^2 \kappa^2}}, \qquad
\Phi = \alpha \phi,
\ee
we obtain (up to boundary terms) the $S_A$ action in (\ref{eq:ourLag}), with an additional $\chi$ field:
\be
S_{\text{Lif-CS}}&\stackrel{k_0^2=0}{=}&S_A+
\dot \chi \psi^\dagger \psi + \nabla_i \chi   \left( i \gamma \psi^\dagger \lrnab^i \psi - \gamma (\nabla^i \chi + 2\alpha \epsilon^{ij} \nabla^j \phi) \psi^\dagger \psi \right) .
\ee
Variation of the action by $\delta \chi$ just gives current conservation. Even more simply, since this action is invariant under the gauge transformation 
\be
\psi&\rightarrow& e^{i \theta(t,x,y)} \psi, \qquad \chi \rightarrow \chi  - \theta(t,x,y),
\ee 
the $\chi$ field is pure gauge and can be removed by performing the above transformation with $\theta = \chi$.  

While this relation to a (modified, gapless) Chern-Simons theory is interesting, we emphasize that the nature of the anyonic statistics is dramatically different in these
theories and in the standard topological Chern-Simons theory.  Notably, marginal parameters of the fixed-point theory enter in determining the anyonic phases.
In terms of the Lifshitz-Chern-Simons parameters,  the anyonic phase $\tha = g \alpha$ takes the form
\be
\tha = \left( 2 \kappa + \frac{f^2}{2 c^2 \kappa} \right)^{-1} .
\ee
That is, at small $c \kappa/f$, $\tha$ is approximately proportional to the Chern-Simons level $\kappa$, rather than inversely proportional to it as was the case when $c^2k_0^2>0$.
This underscores the fact that critical Lifshitz-Chern-Simons theory is very different from Chern-Simons theory itself.  When the $ E^2$ deformation is present, it completely changes the IR physics and, in turn, the way that anyonic statistics arise in the IR.  To see this more explicitly, one can perform the same procedure as above with $c^2k_0^2>0$, in which case one finds that the $\Phi$ kinetic term contains
\be
\LL_{\text{non-critical}} &\sim& \frac{ \nabla^2}{\nabla^2 -k_0^2} \dot \Phi^2
\ee
and thus at momenta less than $k_0$, this would-be kinetic term for $\Phi$ field becomes irrelevant and the $\phi$ action in (\ref{eq:ourLag}) no longer provides a reliable local description of the physics.  

\begin{figure}[ht]
\begin{center}
\includegraphics[width=2.0in]{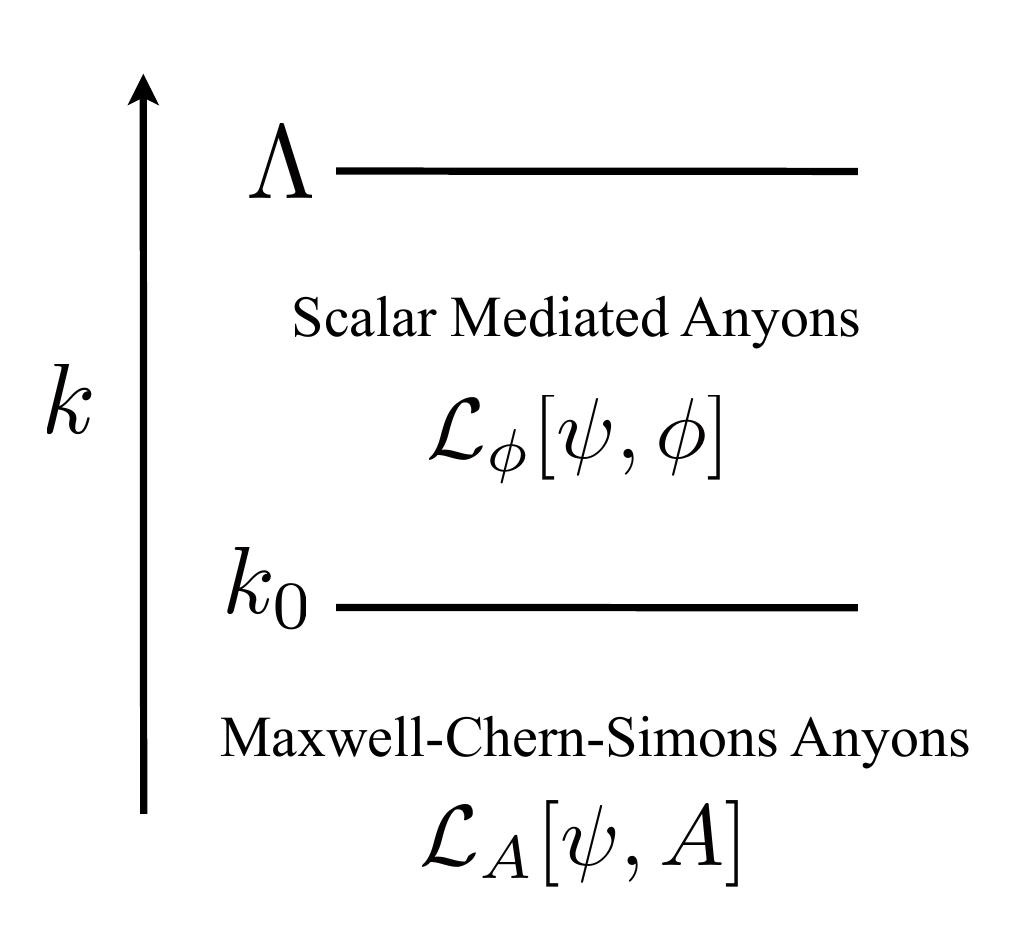}
\caption{At distances shorter than $k_0^{-1}$, Lifshitz Chern-Simons has a description in terms of our local, renormalizable scalar field theory.  There are also local deformations of the scalar theory that correspond to non-local modifications of the Lifshitz Chern-Simons theory. \label{Fig: Scales}}
\end{center}
\end{figure}

We should also stress that under the `Mulligan Duality' between our scalar + fermion theory and the Lifshitz-Chern-Simons theory, local operators in one theory map to non-local operators in the other (as is evident from the formula for $E^i$ in (\ref{cov})).  If one is to imagine obtaining our theory by tuning a scalar + fermion system to criticality, for instance, it is most natural to imagine a space of couplings 
allowing for the presence of the $\phi^2$ and $(\nabla\phi)^2$ operators which are absent at the critical point.  Neither of these has a purely local description in the
Lifshitz-Chern-Simons theory.  Similarly, the duality between local theories that we exhibited holds only in absence of the natural interaction 
\begin{equation}
\label{anotherint}
\delta S=\int dt d^2x ~(\nabla^iE_i) \psi^\dagger \psi
\end{equation}
which is another allowed marginal coupling of the Lifshitz-Chern-Simons theory to the fermions.  Inclusion of (\ref{anotherint}) would render the $\phi, \psi$ description non-local.
The same remark applies to the $E^4$ term which played an important role in \cite{Mulligan:2010wj}.

\section{A Quantum Theory of Anyons}
\label{sec:QuantumAnyons}

In the previous section we showed that the interactions of $\psi$ particles with the $\phi$ field lead to anyonic statistics for the $\psi$s.  Unlike in other calculable field theories of anyons, which are all essentially based on the Chern-Simons action, our anyons arise from the exchange of a dynamical degree of freedom.  Thus it is natural to wonder how the story changes when we include quantum effects.  In particular, one might expect that the couplings evolve logarithmically due to the RG, and that in particular the product $\alpha g$, which sets the anyonic phase, will be scale-dependent.  In fact, we will see that both $\alpha$ and $g$ are invariant to all orders in perturbation theory, so the anyonic phase is completely scale-independent.  In Chern-Simons theory this result followed from topology; in our theory it follows from a combination of symmetries and non-relativistic
kinematics.

We will find it convenient to perform the corresponding calculations using the Wick rotated (i.e. Euclidean) version of the action:
\be
\label{actionis}
S_E &=& \int dt d^2x 
\left[ - \psi^\dag  \partial_t  \psi + \gamma \left| \left(\nabla_i + i\alpha \epsilon_{ij} \nabla^j \phi \right) \psi \right|^2 - g \nabla^2 \phi \psi^\dag \psi 
+   \frac{1}{2}  \left( \dot \phi^2 +  (\nabla^2 \phi)^2  \right)\right]~.
\label{eq:ourLagEuc}
\ee 
We will treat the couplings $ \alpha, \gamma\alpha, g$ in (\ref{actionis}) as small parameters in a perturbative expansion.  (The appearance of $\alpha$ as a small parameter controlling some low-energy couplings, without additional factors of $\gamma$,
will be explained in section \ref{sec:Radiation}).

\begin{figure}[ht]
\begin{center}
\begin{minipage}{0.22\textwidth}\vspace{0.3cm}\includegraphics[width=\textwidth]{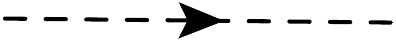}\end{minipage}
\begin{minipage}{0.25\textwidth}
\be
= \frac{1}{\omega^2 + k^4} \nn
\ee
\end{minipage}
\begin{minipage}{0.22\textwidth}\vspace{0.3cm}\includegraphics[width=\textwidth]{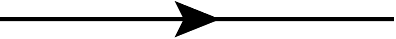}\end{minipage}
\begin{minipage}{0.25\textwidth}
\be
=\frac{-1}{i \omega - \gamma k^2} \nn
\ee
\end{minipage}
\end{center}
\vspace{0.3cm}
\begin{center}
\begin{minipage}{0.22\textwidth}\includegraphics[width=\textwidth]{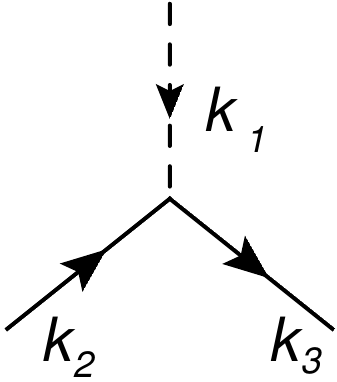}\end{minipage}
\begin{minipage}{0.25\textwidth}
\be
 = -2i \gamma \alpha \epsilon_{ij} k_2^i k_3^j -  g k_1^2 \nn
\ee
\end{minipage}
\begin{minipage}{0.22\textwidth}\includegraphics[width=\textwidth]{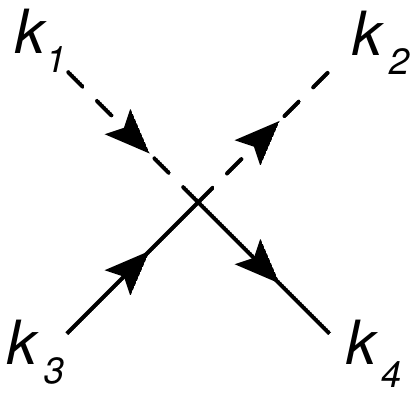}\end{minipage}
\begin{minipage}{0.25\textwidth}
\be
 =-2 \gamma \alpha^2 k_1 \cdot k_2\nn
\ee
\end{minipage}
\end{center}
\caption{These are the Feynman rules for the Lagrangian in  (\ref{eq:ourLagEuc}).  Note that the anyon propagators only have poles on one side of the energy axis, because there are no anti-particles.  }  
\label{fig:FeynmanRules}
\end{figure}

The Feynman rules for this action are summarized in Fig. \ref{fig:FeynmanRules}. The fermions are one component Grassmann fields.   By symmetries and dimensional counting, a four-Fermi interaction is allowed and would be marginal; however, because of the Fermi statistics of $\psi$, this interaction exactly vanishes.  In App.~\ref{sec:AddingSpin} non-relativistic spinors are reviewed and a theory with multiple fermions is formulated and studied;  in this theory there exists a four-Fermi interaction.

The fermion field $\psi$ is non-relativistic and as a consequence it contains particles only -- $\psi$ does not create or destroy antiparticles.  The $\psi$ field can be expanded in terms of normal modes as
\begin{eqnarray}
\psi(x) = \int \frac{ d^2k}{(2\pi)^2} b_{\vec{k}} \exp(-i \omega_{\vec{k}} t + i   \vec{k}\cdot \vec{x})
\end{eqnarray}
with the frequency satisfying the $\psi$ dispersion relation 
\begin{eqnarray}
\omega_{\vec{k}} = \gamma k^2 .
\end{eqnarray}
The creation and annihilation operators satisfy canonical anti-commutation relations
\begin{eqnarray}
\{ b_{\vec{k}}^\dagger, b_{\vec{k}'}\} =  (2\pi)^2\delta^2(\vec{k} - \vec{k}').
\end{eqnarray}
The absence of antiparticles has major implications for the radiative structure of the theory.  The first is that the leading order Wick-rotated propagator of $\psi$ is
\begin{eqnarray}
\Delta_\psi(\omega, k) = \frac{-1}{i \omega - \gamma k^2} .
\end{eqnarray}
This means that all poles for fermions will fall in the upper half-plane, so the Feynman propagator is the retarded propagator.    Therefore, any loop diagram only involving $\psi$ propagators will have poles on only one side of the integration region.  Then, the contour integral can be closed on the other side, and will vanish.  Physically this makes sense because because $\psi$ particle number is conserved and with the absence of anti-particles, there is no way of pair creating fermions.  Another way of seeing this is that a closed fermion loop requires a fermion coming back to the same point in space-time where it was created, thus traveling backwards in time, but since the Feynman propagator is the retarded propagator, this is not possible. The absence of closed fermion loops is an all-orders statement and prevents the $\phi$ propagator from ever being renormalized.

The RG invariance of both $\alpha$ and $g$ depends on a dynamical fact. Consider the following terms in the Lagrangian:
\begin{eqnarray}
\LL_{\rm} &\supset& \gamma \OO_{\gamma_1} + c_{\gamma_2} \OO_{\gamma_2}+ c_{\gamma_3} \OO_{\gamma_3}, \\
\OO_{\gamma_1} &\equiv&  |\nabla^i \psi|^2,\quad \OO_{\gamma_2} \equiv i\ \nabla^i\phi \epsilon_{ij} \psi^\dagger \delfb{}^j \psi, \quad \OO_{\gamma_3} \equiv   (\nabla^i\phi)^2 |\psi|^2.
\end{eqnarray}
In the absence of symmetry assumptions, the above operator coefficients are a priori independent.  
However, the Cauchy-Riemann symmetry ensures through a Ward identity that the coefficients satisfy $c_{\gamma_2} = \alpha \gamma$, $c_{\gamma_3} = \alpha^2 \gamma$, and prevents
$\gamma$, $c_{\gamma_2}$, and $c_{\gamma_3}$ from being separately renormalized.  This fact combined with the absence of wave function renormalization for $\phi$ means that only $\gamma$ is renormalized and not $\alpha$, and their $\beta$ functions are related by $\beta_{\gamma \alpha^2} = \alpha \beta_{\gamma \alpha} = \alpha^2 \beta_{\gamma}$.

The scale invariance of $g$ is more subtle, but it also follows from the symmetry structure and the non-renormalization of $\phi$. The crucial point is that symmetry currents are never renormalized.  So consider the currents in  (\ref{eq:numbercurrent}) and  (\ref{eq:phicurrent}), which are linked because they are both components of the Cauchy-Riemann symmetry current.    They are related by
\be
g (\nabla^i J_{N}^0) =  \nabla^i \nabla^2 \phi  + J^i_{\phi}  + \alpha \epsilon^{ij}  J_{j N}.
\label{eq:PhiCurrent}
\ee
We see that the entire right hand side cannot be renormalized, since loops do not renormalize $\phi$, and the derivative of the current $\nabla^i J_N^0$ must also be RG invariant.  Thus we conclude that $g$ must be invariant as well.  

We will verify that the $\beta$ functions for $\alpha$ and $g$ vanish at one-loop via an explicit calculation later in this section. 
 We will also obtain the RG scaling of the $\gamma$ parameter, confirm the symmetry arguments above at one-loop, and explore the long-distance structure of the theory, studying $\phi$ radiation and the long distance forces between anyons.
While the assumption of the Cauchy-Riemann symmetry connects the coefficients of the operators $\OO_{\gamma_2}$ and $\OO_{\gamma_3}$, it is interesting to ask whether this Cauchy-Riemann symmetry is an attractive or repulsive IR fixed point.   So in what follows we will give these operators different coefficients and explore how they behave when we scale to the IR. 

\subsection{The Definition of Statistics in a Gapless Theory}
\label{sec:RadiationandPotential}
\label{Sec: DefinitionOfStatistics}

Many expositions of anyons (see \cite{Chetan,Wen}) suggest that a gap is necessary for well-defined statistics.  Our theory is gapless.  How are we to be sure that e.g. upon encircling one of our $\psi$ anyons with another, we return to the same
ground state up to a phase (in the Abelian case)?  Couldn't the emission of $\phi$ quanta confound our attempt to measure the statistical phase?

The anyons have several marginal interactions with the massless boson $\phi$.  This means that $\phi$ exchange can give rise to long-range forces, and since $\phi$ number is not conserved, accelerating anyons can radiate soft $\phi$ particles, thereby potentially altering significantly the state of a system of anyons.   One way to state the concern is that when transporting two anyons around each other, it is not possible to return to the exact same state because the Fock space of $\phi$ quanta becomes populated.  One way to avoid this problem is simply to turn on an external potential, such as a harmonic trap.  Two anyons in the ground-state of
the trap will not be able to radiate $\phi$ particles.  One can then ask how the ground state wave function, $\Psi_{GS}(x_1,x_2)$ transforms under the exchange of $x_1$ and $x_2$.  In two spatial dimensions, the
two particle wave function does not have to be single valued and can pick up an anyonic phase.  

Another way of defining the phase is through adiabatic transport.    
For instance, one could measure the anyonic phase with an interference experiment where the two anyons, trapped by an external potential at positions $x_1$ and $x_2$, are interchanged through by going through a path $\CC$ by the
manipulation of the potential:
\begin{eqnarray}
\nonumber
&&|\Psi (0)\rangle= | \psi_1(x_1)\psi_2(x_2) \rangle \stackrel{\CC}{\rightarrow}  \\
&&\qquad\qquad
|\Psi(\CC)\rangle = e^{i\tha+i \vartheta_{\text{M}}(\CC)} ~ r(\CC) | \psi_1(x_1)\psi_2(x_2)\rangle+|\psi_1(x_1)\psi_2(x_2) + \text{multi-part}\rangle.
\end{eqnarray}
Here, $\tha$ is the anyonic phase and $\vartheta_{\text{M}}$ is some path-dependent interacting phase coming from the manipulating potential. 
\footnote{Note that the anyons do not experience a force if one of them is stationary (see (\ref{v-force})), and thus there is no additional phase associated with interactions between anyons.}
$r(\CC) \leq 1$ gives the 
overlap between the initial and final states, and reflects the fact that transport along $\CC$ has entangled the initial state with a population of $\phi$ due to radiation.  
Having $r\ne 1$ is not a problem; but in principle,  IR divergences could have forced $r(\CC)= 0$. However, as we will show below, $r(\CC)$ is IR finite, and can be arbitrarily close to unity in the adiabatic 
limit.

\subsubsection{Radiation}
\label{sec:Radiation}

Given the absence of infrared divergences in the $\psi$ emission of $\phi$ quanta (to be substantiated through
computations below), one can show that the number of $\phi$ quanta radiated in such an adiabatic process is given by
\begin{equation}
\label{NRadiated}
N_{\rm \phi} \sim \omega R^2 f(\omega R^2)
\end{equation}
where $\omega$ is the frequency of the adiabatic transport, $R$ is the distance between the two anyons, and $f$ is non-singular at $0$.   This follows because the $\phi$ radiation rate is proportional to the $\psi$ particle acceleration, although we check it via an explicit computation of the classical radiation rate in App.~\ref{app:adiabatic}.  Taking the limit $\omega R^2 \to 0$ while $R \to \infty$, we are guaranteed that in the limit no $\phi$ quanta will be radiated.  Thus the initial and final state after adiabatic transport can be directly compared, and the anyonic phase is well-defined.\footnote{We expect that similar physics can also explain the robustness of fractional statistics in the presence of soft phonon modes in fractional quantum Hall samples.}

Now let us study soft $\phi$ radiation from accelerating anyons in order to understand the IR structure of the theory at the quantum level.  We will consider an anyon scattering process, and compute the amplitude for radiating a soft $\phi$ boson of momentum $\vec q$.  In the limit of small $\vec q$, the amplitude will take the form
\be
\mathcal{M}_{n_\psi,\phi} \approx \mathcal{M}_{n_\psi} \sum_{a=1}^{n_\psi} \frac{ - 2i \gamma  \alpha \epsilon_{ij} k_a^i q^j - g q^2}{\gamma \left( 2 q \cdot k_a + (\frac{1}{\gamma} + 1)q^2 \right) + i \epsilon },
\ee
because it is  dominated by soft $\phi$ radiation off of the external legs of the scattering process.  We see immediately that the $g$ coupling is sub-leading in the soft limit, so we will drop it in what follows.  Another crucial feature is that in the soft limit, the denominator carries a factor of $\gamma$, which can cancel the $\gamma$ from the $\gamma \alpha$ coupling.  Note that this feature persists as long as $|q| \lsim \gamma |k_a|, |k_a|$.

Keeping only the leading terms, we find that the $\psi$ scattering cross section is
\be
\label{eq:Soft}
\left| \mathcal{M}_{n_\psi} \right|^2 \approx 4\alpha^2 \int \frac{d^2 q}{(2 \pi)^2} \frac{1}{2q^2}     
\sum_{a,b =1}^{n_\psi} \eta_a \eta_b \frac{  \epsilon_{ij} k_a^i q^j \epsilon_{kl} k_b^k q^l }{ \left(2 q \cdot k_a + (\frac{1}{\gamma} + 1)q^2 + i \epsilon \right) \left(2 q \cdot k_b +  (\frac{1}{\gamma} + 1)q^2 - i \epsilon \right) } ,
\ee
where the factor of $\frac{1}{2E_\phi} = \frac{1}{2q^2}$ comes from the definition of the $\phi$ phase space, and the parameter $\eta_a$ is $1$ if particle $a$ is incoming and $-1$ if it is outgoing.  The overall phase-space integrated soft factor is dimensionless, as we should expect, since the emission of soft radiation cannot change the engineering dimension of the hard particle cross section.  

\begin{figure}[ht]
\begin{center}
\includegraphics[width=3.0in]{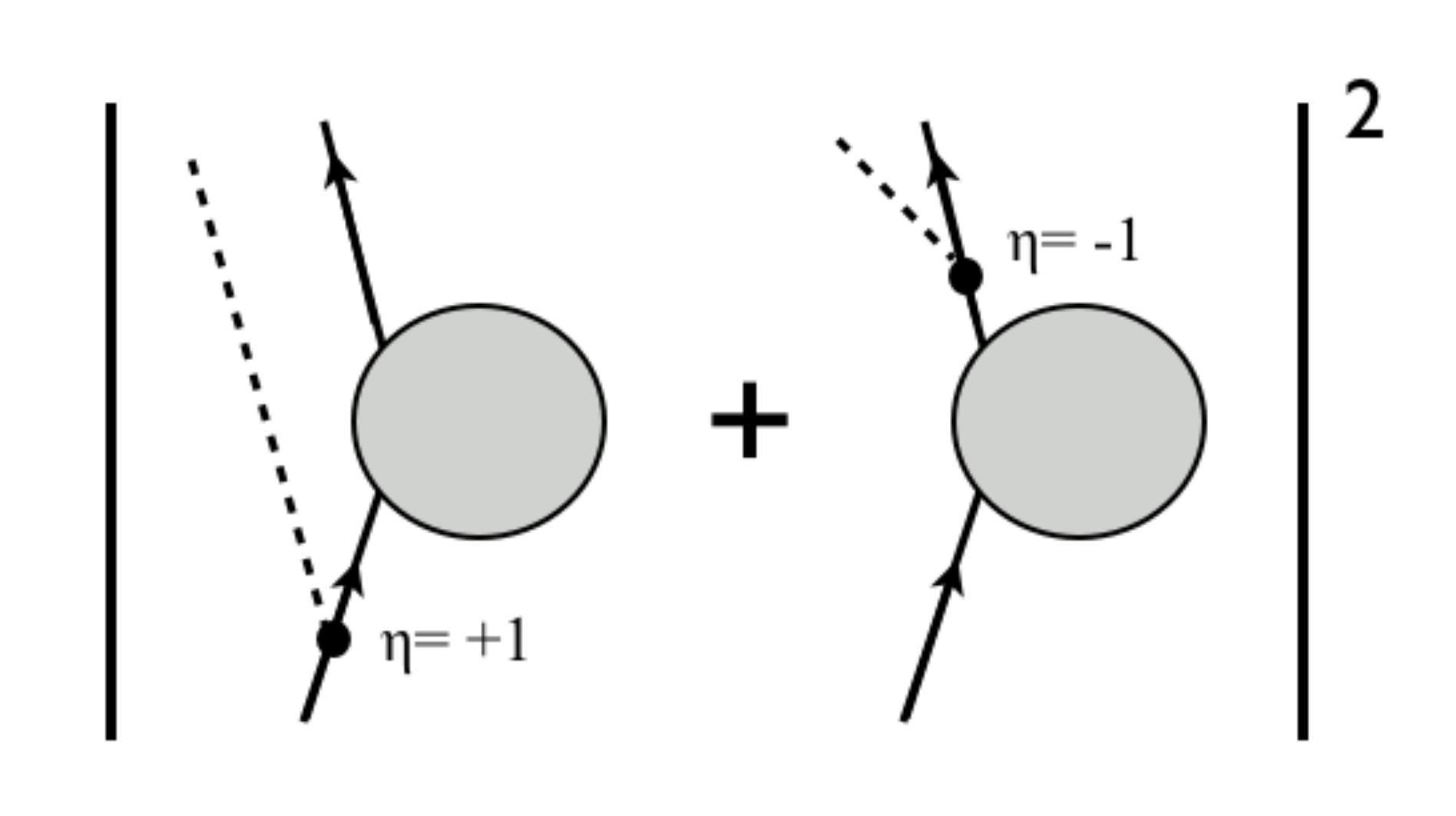}
\caption{Radiation of $\phi$ quanta off of initial and final states cancels in the limit of soft emission, leading to the absence of IR divergences.\label{Fig: Radiation}}
\end{center}
\end{figure}

In the extreme IR region where $|q| \ll \gamma |k_a|, |k_a|$, the soft factor depends only on the directions, and not the magnitudes, of the hard momenta $k_a$, because the $q^2$ factors are negligible in this regime.  If we only consider this region of the integral, then each term in the sum in   (\ref{eq:Soft}) will have a logarithmic IR divergence.  However, an explicit calculation shows that these divergences are completely independent of the momenta $k_a$, and so when we sum over the external legs $a,b$ these IR divergences will cancel between initial and final states.  The calculation is simple when we write $\vec q$ in polar coordinates; it then depends only on the relative angle between $k_a$ and $k_b$, and one can immediately perform the integral and find a result independent of $k_a$ and $k_b$.  Thus $\phi$ emission will be IR finite at one-loop.  As a soft effect, this result exponentiates to all loop order, so there are no IR divergences due to $\phi$ radiation in the theory.

Eq.~(\ref{eq:Soft}) has another striking feature:  as long as $|q| \lsim \gamma |k_a|$ the relevant coupling is not $\gamma \alpha$, which appears in the Feynman rules, but simply $\alpha$ by itself.  The factor of $\gamma$ has been canceled by the nearly on-shell fermion propagator.  As we will see below, the $\beta$ function for $\gamma$ causes it to run towards $0$ at low energies, so it is very natural to consider theories where $\gamma \ll 1$.  This suggests that for these values of $q$, $\phi$ radiation receives a large enhancement, although for small $\alpha$ the theory is clearly under good control.  We leave exploration of the details of this enhancement to future work.  

\subsubsection{Anyon Potential}

Let us begin by noting that there is no long-distance static potential between anyons.  This follows because the $\alpha$ interactions depend on the anyon momenta, while the $g$ coupling has too many derivatives to give a long-range force.  Furthermore, because of the Pauli exclusion principle, which also applies to anyons, if we have only one species of anyons then even contact interactions vanish.  However, when the anyons are in motion, there is a long-range velocity-dependent interaction due to the exchange of $\phi$ bosons
\be
\label{v-force}
V(p,r)=(\gamma \alpha)^2 \int d^2 q \frac{ \epsilon^{ij} p_{1i} q_j  \epsilon^{ab} p_{2a} q_b }{q^4-\gamma^2 (q\cdot (p_1+p_2))^2 } e^{i \vec q \cdot \vec r},
\ee
where the anyons have momenta $\vec p_1$ and $\vec p_2$.  The spatial derivative of this interaction gives a force that depends in a non-trivial way on the anyon momenta.  It is reminiscent of the magnetic force between current-carrying wires.  This potential manifestly vanishes when either of the anyons is motionless; the leading contribution to the force at small $p r$ is
\be
F^i  \sim -(\gamma\alpha)^2 \frac{p_1^j p_2^k }{r}  \left(\delta^{jk} \hat{r}^i +\delta^{ik} \hat{r}^j +  \delta^{ij} \hat{r}^k  -2 \hat{r}^i \hat{r}^j \hat{r}^k \right).
\ee

There are loop induced interactions for $\psi-\psi$ states, and these are marginal interactions.  Naively the ultraviolet behavior of these diagrams will give rise to a local counter-term; however, this vanishes due to Fermi-statistics.  At the diagrammatic level, this follows because there are $t$ and $u$ channel diagrams and the UV divergences cancel between the two channels.  This does not mean that the loops do not give rise to interactions at long distances.   By power counting, the most IR divergent term could be proportional to $\log q^2$ (where $\vec q$ is the momentum exchange) and would give rise to a $ 1/r^{2}$ potential. However, this would-be IR divergence is related to the soft limit of $\phi$ emission by the optical theorem, and since soft emission is actually IR finite, these loop effects must be as well.  This means that $\phi$ loops between anyons should only generate positive powers of $q^2$, which do not give rise to long range forces.

\subsection{Propagator Corrections}

\begin{figure}[ht]
\begin{center}
\begin{minipage}{0.35\textwidth}\vspace{0.3cm}\includegraphics[width=\textwidth]{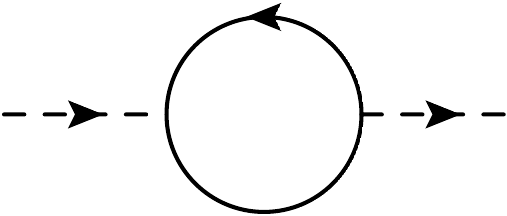}\end{minipage}
\begin{minipage}{0.08\textwidth}$\phantom{s}$\end{minipage}
\begin{minipage}{0.35\textwidth}\vspace{0.3cm}\includegraphics[width=\textwidth]{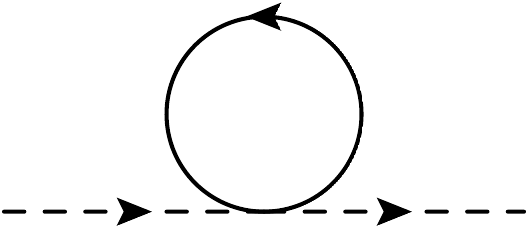}\vspace{1cm}$\phantom{s}$\end{minipage}
\end{center}
\caption{Both of the above one-loop renormalizations of the boson propagator vanish, because all poles of the fermion propagators are on the same side of the integration contour.  Physically this follows because our non-relativistic theory does not contain anti-anyons, so pair creation is impossible. 
 }
\label{Fig:BosonPropagator}
\end{figure}

As discussed above, the $\phi$ propagator is not renormalized at any order in perturbation theory due to conservation of $\psi$ number and the absence of $\psi$ anti-particles in our non-relativistic theory.  At a computational level, $\phi$ is not renormalized by $\psi$ loops because the resulting energy integrals can be performed by contour integration in a half-plane that contains no poles.  For example, the boson propagator correction in figure \ref{Fig:BosonPropagator} is proportional to 
\be
\INT \frac{f(g, \alpha, \tilde k, k) }{\left( i(\tilde \omega - \omega ) - \gamma (\tilde{k}-k)^2 \right) \left( i \tilde \omega - \gamma \tilde{k}^2 \right)} .
\ee
We can perform the $\tilde \omega$ integral by closing the contour in the upper half-plane, giving a vanishing result.  Another way to view this result is by computing in position space, where the fermion propagators are causal.  In that case fermion loops must vanish because they loop back to their initial time.

\begin{figure}[ht]
\begin{center}
\begin{minipage}{0.35\textwidth}\vspace{0.3cm}\includegraphics[width=\textwidth]{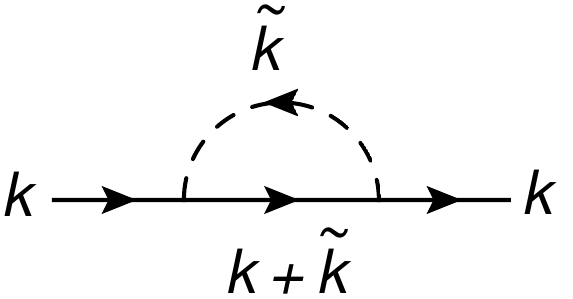}\end{minipage}
\begin{minipage}{0.08\textwidth}$\phantom{s}$\end{minipage}
\begin{minipage}{0.35\textwidth}\vspace{0.3cm}\includegraphics[width=\textwidth]{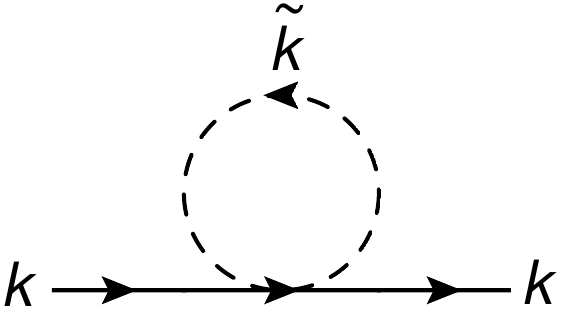}\vspace{1cm}$\phantom{s}$\end{minipage}
\end{center}
\caption{ The left one-loop diagram renormalizes the anyon propagator and leads to a non-trivial $\beta$ function for the $\gamma$ parameter. The seagull diagram on the right has no external momentum flowing through
the loop, and so renormalizes only the chemical potential term, which is tuned to zero.}
\label{Fig:FermionPropagatorAnyons}
\end{figure}

There is a non-vanishing one-loop correction to the $\psi$ propagator that is given by the diagram\footnote{There is also a energy and momentum independent bubble diagram, but this merely contributes a divergent $\psi$ mass, which we are tuning to zero. } in Fig. \ref{Fig:FermionPropagatorAnyons}. Our interest is in the RG flow of the couplings in the theory, so we can neglect any contributions that are not UV divergent.  The corresponding integral is
\begin{eqnarray}
\label{pig}
\delta \Delta^{-1}_\Psi(\omega,k)= \INT 
 \frac{1}{\tilde{\omega}^2 + \tilde{k}^4} \frac{-1}{i(\tilde \omega - \omega ) - \gamma (\tilde{k}-k)^2} 
 \left(g^2 \tilde k^4 - 4\gamma^2 \alpha^2 \epsilon^{ij} \epsilon^{mn} k_i (-\tilde k_j) \tilde k_m (-k_n)  \right) .
\end{eqnarray}
There is also a term proportional to $g \alpha$, but it includes only a single factor of $\epsilon^{ij}$.  Since the result can only depend on $k_i$, by rotational invariance this $g \alpha$ term must vanish, so we have dropped it.

In terms of divergence structure, the lowest order renormalization of the propagator can give rise to a non-zero mass for $\psi$.   The $g^2 \tilde k^4$ term results in the momentum integral being quadratically divergent for the fermion mass.  Due to the $z=2$ scaling, this quadratic divergence in momentum is equivalent to a linear divergence in energy. This is directly analogous to the linear divergence of the self-energy of the electron in non-relativistic QED without antiparticles.   This mass must be fine-tuned to vanish in this theory because of the absence of a chiral symmetry. 
 Similarly, the  $\alpha^2$ seagull correction also gives a similar divergence to the mass.   In contrast, the $\gamma^2 \alpha^2$ correction in (\ref{pig}) (which arises from
 the scalar-fermion interactions in the covariant derivative) does not give a fermion mass.

Both the $g^2$ term and the  $\gamma^2 \alpha^2$ term in (\ref{pig}) give a logarithmically divergent field strength renormalization, while the $\gamma \alpha^2$ seagull loop does not depend on external momenta.   Keeping only the UV divergent pieces and using rotation invariance to set $\tilde k_i \tilde k_j \cong \frac{1}{2} \delta_{ij} \tilde k^2$ inside the integral, the result can be simplified to
\be
\frac{1}{2 (1 + \gamma) }   \int \frac{d^2 \tilde k}{(2 \pi)^2 } \left[ g^2 + \left( 2\gamma^2 \alpha^2 + \frac{ g^2 \gamma  (\gamma -1)}{(1 + \gamma)^2} \right) \frac{   k^2}{ \tilde k^2} +\frac{g^2}{1+ \gamma} \frac{ i\omega}{\tilde k^2}  \right] .
\ee
In addition to a divergent fermion mass, we have found a shift of the $\psi^\dag i \partial_t \psi$ kinetic term from $g$ and of $\psi^\dag \nabla^2 \psi$ from both $\alpha$ and $g$.  
In terms of the renormalized Lagrangian, this means that we obtain a wave-function renormalization counter-term $\delta Z_\psi$ and a counter-term for $\gamma$:
\be
 \delta Z_{\psi} &=&  -\frac{g^2}{2 (1 + \gamma)^2} \int \frac{d^2 \tilde k}{(2 \pi)^2} \frac{1}{\tilde k^2} , \\
\delta_\gamma &=& \frac{1}{2 (1 + \gamma) }   \left[  \left( 2\gamma^2 \alpha^2 + \frac{ g^2 \gamma  (\gamma -1)}{(1 + \gamma)^2} \right)  \right]\int \frac{d^2 \tilde k}{(2 \pi)^2 \tilde{k}^2} .
\ee
It is easy to regulate these divergent integrals with a hard cut-off $\Lambda$ on spatial momentum $\tilde{k}$.\footnote{A hard momentum cut-off regulator technically breaks the Cauchy-Riemann symmetry, but none of the counter-terms using this regulator violate the symmetry.  It is also possible to use Pauli-Villars or possibly dimensional regularization which explicitly preserves the Cauchy-Riemann symmetry.} Putting $\delta_\gamma$ and $\delta Z_\psi$ together, the $\beta$ function for $\gamma$ is simply
\be
\beta_\gamma =  \frac{\partial {\delta_\gamma - \gamma \delta Z_\psi}}{\partial \log \Lambda} = \frac{\gamma^2  \left( g^2+\alpha ^2 (1+\gamma )^2\right)}{2\pi
   (1+\gamma )^3}.
\ee
Thus we see that $\gamma$ runs to zero at low energies; however, it only reaches zero at a asymptotically long wavelengths, running for $\gamma\ll 1$ as
\begin{eqnarray}
\gamma(\Lambda_{\text{IR}}) \simeq\left(\gamma(\Lambda_{\text{UV}})^{-1}+  \frac{(g^2+\alpha^2) \log \Lambda_{\text{UV}}/\Lambda_{\text{IR}}}{2\pi}\right)^{-1}.
\end{eqnarray}
Since $\gamma \propto  m_\psi^{-1}$, this means that $\psi$'s effective mass is becoming larger at long distances.

\subsection{Vertex Corrections}
\label{sec:vertexcorrection}

\begin{figure}[ht]
\begin{center}
\includegraphics[width=0.30\textwidth]{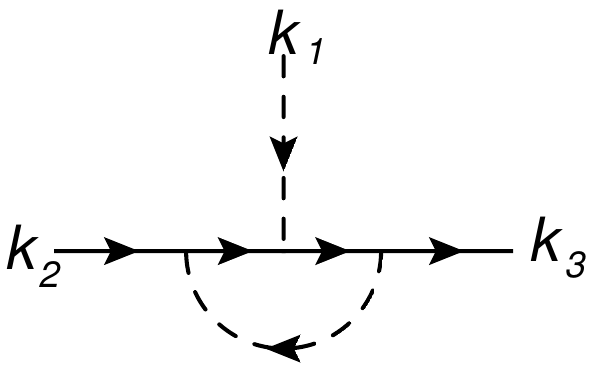}
\end{center}
\begin{center}
\begin{minipage}{0.30\textwidth}\includegraphics[width=\textwidth]{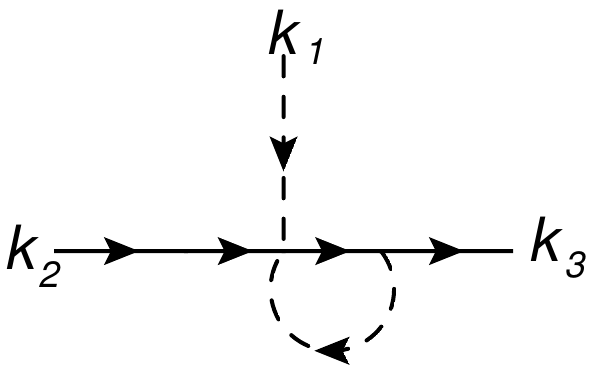} \end{minipage}
\begin{minipage}{0.2\textwidth}\phantom{sssssssssss}\end{minipage}
\begin{minipage}{0.30\textwidth}\includegraphics[width=\textwidth]{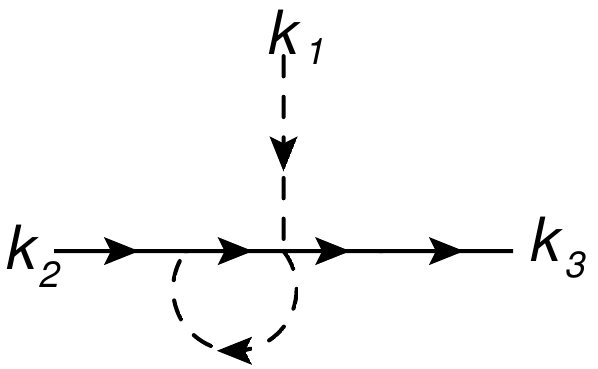} \end{minipage}
\end{center}
\caption{ These diagrams give the one loop renormalization of the 3-pt couplings between the anyons and $\phi$.  However, they do not lead to any RG running beyond that of the parameter $\gamma$, which sets the effective anyon mass.  The $\alpha$ and $g$ couplings are not renormalized.  }
\label{Fig: YukawaRenormalization}
\end{figure}

The cubic vertices in the theory are protected from renormalization as discussed above.  However, it is illuminating to consider what happens when one perturbs away from the symmetric theory.  Among other things, this will allow us to see if the theory naturally flows to the symmetric point.  
One convenient way of parameterizing this perturbation is to modify the coefficient of the $\phi \phi \psi \psi$ quartic interaction:
\be
\LL \supset - 2 \gamma \alpha^2 ({\nabla} \phi)^2 \psi^\dagger \psi \rightarrow - 2\gamma  \tilde{\alpha}^2 ({\nabla} \phi)^2 \psi^\dagger \psi .
\ee
This is clearly equivalent to adding to (\ref{eq:ourLagEuc}) an additional quartic piece
\be
\LL \rightarrow \LL - \half A (\nabla \phi)^2 \psi^\dagger \psi, \qquad A \equiv 4\gamma(\alpha^2 - \tilde{\alpha}^2).
\ee
The $\beta$ functions are calculated in App.~\ref{Sec: Vertex} and are
\be
\beta_{\gamma \alpha} &=& 
\alpha \gamma^2 \frac{ g^2 + \tilde{\alpha}^2 (1+\gamma)^2}{2\pi(1+\gamma)^3},\\ 
\beta_{\gamma \tilde{\alpha}^2} &=& 
 \gamma^2 \frac{ g^2 \alpha^2 + \tilde{\alpha}^4 (1+\gamma)^2}{2\pi(1+\gamma)^3}, \\
\beta_g &=& 
 g \gamma^2 \frac{ \alpha^2 - \tilde{\alpha}^2 }{2\pi(1+\gamma)^2} =  \frac{A g \gamma}{8\pi(1+\gamma)^2} .
\ee

Note first of all that this explicitly verifies that when $A=0$, we have $\beta_g =0$ and $\beta_{\gamma \alpha} = \alpha \beta_\gamma$ and $\beta_{\gamma \tilde{\alpha}^2} = \alpha^2 \beta_{\gamma}$, so neither $g$ nor $\alpha$ run, as expected.  
The $\beta$ function for the coupling $A$, which is a measure of how badly the symmetry is broken, is
\be
\beta_A &=& - \frac{A^2}{8\pi(1+\gamma)}.
\ee
Thus, the symmetric theory $A=0$ is an attractive fixed point if $A(\Lambda_{\text{UV}})<0$, and a repulsive fixed point for $A(\Lambda_{\text{UV}})>0$.  If $A(\Lambda_{\text{UV}})>0$, then the theory will become non-perturbative at a  scale, $\Lambda_{\text{IR}}$ given by
\begin{eqnarray}
\Lambda_{\text{IR}} \simeq \Lambda_{\text{UV}} \exp\left( -\frac{ 8\pi (1+\gamma)}{A(\Lambda_{\text{UV}})}\right).
\end{eqnarray}
  This means that our theory with the Cauchy-Riemann symmetry is an attractive IR fixed point for half of the naive UV parameter space.

\section{Generalization to Non-Abelian Anyons}
\label{sec:NonAbelian}

The Abelian model we have presented above gives rise to anyons, but we can produce a semi-classical theory of `non-Abelian anyons' as well by promoting $\phi$ to the adjoint representation of a Lie group.  Performing this promotion in the most obvious way gives the action
\be
\label{eq:NonAbelianLinear}
S=\int dt d^2x \left[ \frac{1}{2} \left( (\dot \phi^A)^2  -  \left( \nabla^2 \phi^A \right)^2 \right)  -\psi^\dag_a  i\partial_t  \psi^a - \gamma \left| \mathcal{D}_{i} \psi^a \right|^2 + g  \psi^\dag_a \left(\nabla^2 \phi^A T^A{}^a_b \right) \psi^b  \right],
\ee
where the covariant derivative is $\mathcal{D}^i{}^a_b = \nabla^i \delta^a_b  + i \alpha  \epsilon^{ij}\nabla_j \phi^A T^A{}^a_b$.
Static $\psi$ particles source the $\phi^A$ as
\be
\nabla^2 \phi^A = g \rho^A \equiv g \psi^\dag_a T^A{}^a_b \psi^b.
\ee
The first quantized action for $\psi$ involves the matrix  $\phi^A T^A$, and thus it rotates the $\psi^a$ vector by 
\be
\tha^A = g \alpha \int_M d^2 x \ \!  \rho^A (x) 
\ee
when a $\psi$ particle encircles non-trivial charge.  This results in a $U(N)$ rotation  in the flavor space 
\begin{eqnarray}
U^A_B = \exp(i \tha^AT^A) .
\end{eqnarray}
Note that this matrix involves a sum over all the charges enclosed by the path.  

As an illustrative example, let us take $\phi$ to be a $U(2)$ valued matrix with $\psi$ in the fundamental representation.  If we move an anyon, $\psi_0$, along a path enclosing two other anyons in the state
\be
\psi_1^a\propto \left( \begin{array}{c}
1\\ 
0\end{array} \right)
\ \ \ \mathrm{and} \ \ \ 
\psi_2^a\propto \frac{1}{\sqrt{2}} \left( \begin{array}{c}
1\\ 
1\end{array} \right),
\ee
then the encircling anyon picks up a matrix valued phase given by the integral of $\rho_A(x)$ within its path.   In this example $\rho_A$ will include two delta functions at the locations of $\psi_1$ and $\psi_2$, and so the phase will be $\psi_1 \psi_1^\dag + \psi_2 \psi_2^\dag$.  This rotates the encircling anyon by 
\be
\psi_0 \to \exp \left[ \frac{i}{2} \alpha g \left( \begin{array}{cc}
3 & 1\\ 
1& 1
\end{array} \right) \right] \psi_0.
\ee
In general, we can decompose the representation of a sum of charges into the irreducible representations contained in their tensor product; Clebsch-Gordon coefficients determine the action of these representations on an encircling anyon.  

At a full quantum level the action in  (\ref{eq:NonAbelianLinear}) will be renormalized, but as written it lacks the symmetries necessary to protect the anyonic phase from renormalization.  One could consider trying to extend the relationship of the scalar theory to Lifshitz-Chern-Simons theory, detailed in \S\ref{sec:ComparisonChernSimons}, to the non-Abelian theory, but
the renormalization of the non-Abelian Lifshitz-Chern-Simons theory is in itself a challenging problem.
We will leave a detailed analysis to the future, but will here simply point out that we can construct a more symmetric action by using an extreme limit of the methods discussed in  \S\ref{sec:ComparisonChernSimons}.   

In the Abelian case we can begin with a theory of $\phi$ and $\psi$ fields coupled to a Chern-Simons field via
\be
\int dt d^2 x  \left[ -\psi^\dag i D_t \psi - \gamma_\psi | D_i \psi |^2 +  \frac{1}{2} \dot \phi^2 - \frac{1}{2} \left( \nabla^2 \phi \right)^2 + g \nabla^2 \phi \psi^\dag \psi + \alpha \kappa E_i {\partial}^i \phi +  \kappa \epsilon^{\mu \nu \rho} A_\mu d_\nu A_\rho \right],
\ee
The covariant derivative $D_\mu = \partial_\mu + i A_\mu$ and note that we have included a kinetic mixing between $\phi$ and $A_\mu$ which grows with $\kappa$.  If we integrate out $A_\mu$ and take the limit of large Chern-Simons level, so that $\kappa \to \infty$, we (somewhat miraculously) recover our local Abelian theory of $\phi$ and $\psi$ particles.  

This procedure has a natural generalization to the non-Abelian case.  We begin with a non-Abelian Chern-Simons action and introduce a kinetic mixing between $A_\mu$ and $\phi$, giving
\be
\label{eq:NonAbelianNonLinear}
S &=& \int dt d^2x \left[ \frac{1}{2} \left( (\mathcal{D}_t \phi_{a})^2  -  \left( \mathcal{D}_i^2 \phi_{a} \right)^2 \right)  - \psi^\dag_A  i \mathcal{D}_t  \psi^A - \gamma \left| \mathcal{D}_{i } \psi_A \right|^2 + g  \psi^\dag_A \left(\mathcal{D}_i^2 \phi^a T_a^{AB} \right) \psi_B \right.
\nonumber \\
&& + \left. \kappa \epsilon^{\mu \nu \rho} \Tr \left[ A_\mu \mathcal{D}_\nu A_\rho \right] + \kappa \alpha  \Tr \left[ E_i \mathcal{D}^i \phi \right]  \right],
\ee
where all derivatives are gauge covariant and $E_i$ is the electric field constructed from the non-abelian Chern-Simons gauge field.
When we integrate out $A_\mu$ in $A_0 = 0$ gauge and take the limit of large Chern-Simons level and large kinetic mixing, so that $\kappa \to \infty$, we obtain an interacting theory of $\phi_{AB}$ and $\psi_A$ with a non-linear Cauchy-Riemann type symmetry.  Viewed as a matrix, the solution for $A_i$ is
\be
A_i = \frac{1}{1 + i \alpha [ \phi,  \ ] } \epsilon_{ij} \nabla^j \phi  + \mathcal{O} \left(\frac{1}{\kappa} \right),
\ee
which is formally defined by its series expansion in the commutator.  This theory is local in derivatives, but to see the full symmetry structure one must work to all orders in the dimensionless $\phi_{A}$ field.  It will be interesting to study this theory or its critical Lifshitz-Chern-Simons equivalent in the future.

There is an extensive literature on non-Abelian anyons, starting with \cite{Moore}, which largely makes use of the deep connection between Chern-Simons gauge theories and rational conformal field theories.   It would be interesting to explore the relationship between this theory and the approaches based on rational conformal field theory.

\section{Discussion}
\label{sec:Discussion}

We have presented a novel field theory in $2$+$1$ dimensions describing the marginal couplings of $\psi$ and $\phi$ particles with an $\omega \propto k^2$ dispersion relation.  The $\psi$ interactions with the propagating $\phi$ field lead to an anyonic phase when one $\psi$ particle encircles another, and the associated couplings are exactly protected from renormalization by a combination of symmetries and dynamics.  These facts have been checked at one-loop by computing the renormalization group scaling of the theory, and we have also found that at low energies, the effective mass of the $\psi$ particles grows logarithmically, while the four-anyon contact interaction becomes strongly attractive.  The crucial Cauchy-Riemann symmetry was also found to be attractive under renormalization group flow in half of the UV parameter space.  In the future it will be important to investigate the theory at finite density and to understand whether it can be constructed in the laboratory.   We will close with some naive comments and speculations concerning these issues.

As a starting point, one can ask what should be the interpretation of the bosonic field $\phi$?  We leave the question of a realistic implementation of this model to future work, however let us briefly comment on one direction.  It is noteworthy that two of the symmetries which $\phi$ enjoys are $\phi \rightarrow \phi + c$ and $\phi \rightarrow \phi + v t$,  where $c$ and $v$ are constant. These are the same as the transformations of a space coordinate under translations and Galilean boosts.    It is therefore natural to conjecture that $\phi$ should describe the fluctuating
height of a membrane on which the $\psi$ excitations reside.  With such an interpretation a coupling such as $(\nabla \phi)^2 \psi^\dagger\psi$ would occur automatically.  The action 
\be
S= \int dtd^2x \left[ \dot \phi^2 - T (\nabla \phi)^2 - \kappa (\nabla^2\phi)^2 \right]
\ee
would describe the fluctuations of a height field itself.  Here $T$ is the membrane tension and $\kappa$ is related to the extrinsic curvature.  Having a completely tensionless membrane is likely unstable, however it may be possible to have a hierarchy of scales
between the membrane tension and the cutoff of the theory ($T \ll \Lambda^2$).  So far the theory has a parity symmetry in the direction perpendicular to the membrane ($\phi \rightarrow -\phi$).  However, once such a parity is broken by, for example, background fields which also break T-symmetry, terms such as our $g$ and $\alpha$ couplings to $\psi$ will be generated.

Fermions with an $\omega \propto k^2$ dispersion relation can arise in a variety of ways, for example as non-relativistic particles in a trap or through the phenomenon of quadratic band touching \cite{QBT}.  In the latter case, there would be excitations representing both particles and holes, so that in the language of our theory, there would be both $\psi$ particles and anti-particles.   This would alter our theory in an interesting qualitative way, because it would introduce non-vanishing fermionic loop corrections.  Our theory would be similarly affected by a large $\psi$ density (chemical potential).  Thus it will be important in the future to understand how the theory behaves when the $\psi$ particles have a less trivial vacuum structure.

The connection with Lifshitz-Chern-Simons theory suggests another possible avenue for experimental realization.  The theory of \cite{Mulligan:2010wj} was proposed in part
as a potential description of a critical point governing a phase transition between conventional fractional quantum Hall states and fractional quantized Hall nematic phases, seen in a recent
experiment on the $\nu=7/3$ fractional quantum Hall plateau \cite{Eisenstein}.  The considerations in \S2.1\ suggest that anyonic quasiparticles in the fractional quantum Hall
phase could have modified behavior as one approaches the critical point, with a measurable change in their anyonic phases.  However, the coefficient of the $B^2$ term which
modifies the statistics could be naturally small (as the operator is irrelevant away from the fixed point), and it is possible that rounding
of the transition would render such effects un-observable.

Finally, it will be interesting to understand global issues surrounding the anyonic phase.  In Chern-Simons theories, various arguments suggest that the anyonic phase must be quantized.  However, it appears that these arguments  for quantization \cite{Wen} break down when applied to our theory, due to the absence of a gap and subtle differences between a Chern-Simons gauge field and our propagating scalar field.  It will be important to study these issues further, especially since the behavior of physical systems of anyons changes dramatically depending on the value of the anyonic phase and the filling fraction of anyonic Landau levels.

\section*{Acknowledgments}

We would like to thank Maissam Barkeshli, Steve Kivelson, Sri Raghu, and Oskar Vafek for very helpful discussions, and John McGreevy and Mike Mulligan for many useful comments on an early draft.  We are particularly grateful to Mike Mulligan for educating us about the content of \S2.1.  ALF was partially supported by ERC grant BSMOXFORD no. 228169. SK is supported in part by the NSF under grant no. PHY-0756174. SK, JK and JW acknowledge support from the US DOE under contract no. DE-AC02-76SF00515. EK is supported by DOE grant DE-FG02-01ER-40676, NSF CAREER grant PHY-0645456.

\appendix

\section{Derivation of the First Quantized Action}
\label{FirstQuantized}

We can derive the first-quantized action for a $\psi$ particle by first deriving the appropriate Hamiltonian for a one $\psi$ state, and then by matching a first-quantized Lagrangian to this Hamiltonian.  We define the canonical momentum for the $\psi$ field as
\be
\Pi_\psi = \frac{\delta \mathcal{L}}{\delta \dot \psi} = -i \psi^\dag ,
\ee
and then compute the Hamiltonian $\Pi_\psi \dot \psi - \mathcal{L}$ for $\psi$ particles, which is
\be
H_\psi = \int d^2 x \ \! \psi^\dag \left(   -\gamma \left(\nabla_i + i\alpha \epsilon_{ij} \nabla^j \phi \right)^2 - g \nabla^2 \phi   \right) \psi.
\ee
Taking the matrix element of this with a one-particle $\psi$ state with momentum $\vec p$, we find
\be
H_\psi(p) =  \gamma \left(p_i + \alpha \epsilon_{ij} \nabla^j \phi \right)^2 - g \nabla^2 \phi   
\ee
Now let us check that the first quantized action
 \be
S_\psi =  \int dt \left[ \frac{\dot x^2}{4 \gamma} - \alpha \dot x_i \epsilon^{ij} \nabla_j \phi(\vec x) + g \nabla^2 \phi(\vec x) \right]
 \ee
 matches with the Hamiltonian we have just derived from the field theory.  Quantizing this action we see that
 \be
 p_i = \frac{\delta L}{\delta \dot x_i} = \frac{\dot x}{2 \gamma} -  \alpha \epsilon_{ij} \nabla^j \phi
 \ee
So the associated first-quantized Hamiltonian is 
\be
 p_i \dot x_i - L &=& 2 \gamma  \left( p_i +   \alpha \epsilon_{ij} \nabla^j \phi  \right) \left(p_i - \frac{1}{2} \left( p_i +  \alpha \epsilon_{ij} \nabla^j \phi \right) +   \alpha \epsilon_{ij} \nabla^j \phi   \right) - g \nabla^2 \phi 
 \\
 &=&  \gamma \left(p_i + \alpha \epsilon_{ij} \nabla^j \phi \right)^2 - g \nabla^2 \phi   
\ee
where in the first line we have written $\dot x_i$ in terms of $p_i$ and factorized $\dot x_i$ out for convenience.  If we associated the canonical momentum $p_i$ from the first-quantized action with the particle momentum $p_i$ of the second quantized action, we find exact agreement between their Hamiltonians, as desired.  So we have the correct first-quantized action for the $\psi$ particles, and we can view the $\psi$ particle mass as 
\be
m_\psi = \frac{1}{2 \gamma}
\ee
Finally, we can derive the equations of motion for a single $\psi$ particle.  Varying with respect to $\vec x(t)$, we find
\be
\frac{1}{2\gamma} \ddot x_i - \alpha \epsilon_{ij} \nabla^j \dot \phi + \alpha \dot x_j \epsilon^{jk} \nabla_i \nabla_k \phi - g \nabla_i \nabla^2 \phi = 0
\ee
If we interpret $A_i = \epsilon_{ij} \nabla^j \phi$ and set $A_0 = 0$ as though by gauge choice, then this equation can be rewritten as
\be
\frac{1}{2\gamma} \ddot x_i =  \alpha E_i + \alpha \epsilon_{ij} \dot x^j B + g \nabla_i B 
\ee
where the magnetic and electric fields are $B = \epsilon^{ij} \nabla_i A_j$ and $E_i = \dot A_i$.  The first two terms are the standard Lorentz force, while the last term is a more unusual result of the direct coupling of $\psi$ density to $\nabla^2 \phi$.

\section{Soft Radiation from Adiabatic Orbits}
\label{app:adiabatic}

As we move one $\psi$ particle around another one, $\phi$'s will be radiated due to the acceleration of the $\psi$.  One might worry that even in the limit where the $\psi$ orbit is taken to be extremely large (i.e. large radius $R$) and the velocity extremely slow (i.e. small frequency $\omega$), some finite number of $\phi$'s could still be radiated, preventing a sharp comparison between the initial state of just two $\psi$'s before the orbit and the final state after the orbit is completed.  To compute the emission of $\phi$ radiation due to oscillatory $\psi$ motion, we begin with the equation of motion for $\phi$:
\be
\ddot{\phi} + \nabla^4 \phi &=& g \nabla^2 \rho - \alpha \epsilon^{ij} \nabla^i J_N^j .
\ee
The density piece is suppressed by $\nabla^2$, which causes it to be subleading  in the soft limit, so we will just consider the leading contribution, from $J_N^i$.  Since a general source can be decomposed into Fourier modes, one may without loss of generality take the time-dependence of the current to be a single frequency:
\be
\vec{J}_N(x,t) &=& \vec{J}_N(x) e^{i \omega t},
\ee
and similarly for $\phi(x,t)$.  
Ultimately, we care only about the parametric dependence of $\phi$ radiation that escapes to spatial infinity, so we will solve for its behavior in the limit where the size $R$ of the source satisfies $\omega R^2 \ll 1$, and drop numeric factors.
By the method of Green's functions, at small $\omega R^2$ and large $\omega r^2$, we obtain the following multipole expansion for $\phi$:
\be
\phi(x) &\sim& \frac{e^{-i \sqrt{\omega} r }}{\omega (\omega r^2)^{\frac{1}{4}}} \sum_{n=0}^\infty \frac{(-i \sqrt{\omega})^n}{n!} \int d^2 y (\hat{x} \cdot y)^n \alpha \epsilon^{ij} \nabla^i J_N^j(y)  .
\label{eq:multipole2}
\ee
In the adiabatic orbit of one $\psi$ around another, the total charge inside a large spatial region is not oscillating, so the leading contribution will arise from the dipole moment. Let us focus on a 
simple description of the current $\vec{J}$ for the $\psi$ particle (which is moving in a circular orbit with radius $R$ and velocity $v=\omega R$) that is adequate for obtaining the parametric dependence of $\phi$ at large distances:
\be
\vec{J}_N(x) &\sim& \omega |x| f(|x|/R) \hat{\theta},
\ee
where $f(|x|/R) \sim R^{-2} \exp (-x^2/R^2)$ is a normalized distribution centered at $|x|=0$.  
Plugging this into  (\ref{eq:multipole2}), the dipole ($n=1$) contribution to $\phi$ is
\be
\phi(x) &\sim& \alpha \sqrt{ \omega R^2} \frac{e^{-i \sqrt{\omega} r}}{(\omega r^2)^{\frac{1}{4}}}.
\ee
Now, we wish to compute the number of $\phi$'s radiated, which is given by the $\phi$ flux through a sphere of size $r$.  The energy flux is $T^{0i}$:
\be
T^{0i} &\sim& \dot{\phi} \nabla^i \nabla^2 \phi \sim \omega \phi \nabla^i \nabla^2 \phi ,
\ee
so the number flux is $\omega^{-1} T^{0i} \sim  \phi \nabla^i \nabla^2 \phi$. 
Integrating this over a surface at fixed $r$ gives
\be
\dot{N}_\phi \sim r \int  d \theta \left[  \phi \vec{\nabla} \nabla^2 \phi \right] \cdot  \hat{r} \sim   \alpha^2 \omega^2 R^2.
\ee
Thus, the total number of $\phi$'s that are radiated over a single period $\Delta t \sim \omega^{-1}$ at small $\omega R^2$ is parametrically
\be
N_\phi \sim  \alpha^2 \omega R^2 ,
\ee
and therefore, by taking a limit $\omega \rightarrow 0, R\rightarrow \infty$ such that $\omega R^2 \rightarrow 0$,  the total number of radiated $\phi$ particles can be made arbitrarily small.

\section{Cubic Vertex Corrections}
\label{Sec: Vertex}

Here, we give a derivation of the $\beta$ functions of the cubic interactions in the theory, which were presented in section  \S\ref{sec:vertexcorrection}. 
First, we want to compute the top diagram in Fig. \ref{Fig: YukawaRenormalization}.  This takes the form
\be
&& \int \frac{d \tilde{\omega}}{2 \pi} \frac{d^2 \tilde{k}}{(2\pi)^2} \left( -2i \gamma \alpha \epsilon^{ij} k_2^i \tilde{k}^j - g \tilde{k}^2 \right) \left( -2i \gamma \alpha \epsilon^{ij} (k_2+ \tilde{k})^i (k_3 + \tilde{k})^j - g k_1^2 \right) \left(- 2i \gamma \alpha \epsilon^{ij} \tilde{k}^i k_3^j - g \tilde{k}^2 \right) \nn\\
&& \ \ \ \ \  \times \frac{-1}{i(\omega_2 +\tilde{\omega}) - \gamma (k_2 + \tilde{k})^2 }  \frac{-1}{i(\omega_3 +\tilde{\omega}) - \gamma (k_3 + \tilde{k})^2 } \frac{1}{\tilde{\omega}^2 + \tilde{k}^4} . 
\ee
Closing the $\tilde{\omega}$ contour in the upper half-plane, we pick up the pole at $\tilde{\omega} = i \tilde{k}^2$. We can Taylor expand the resulting integrand  at large $\tilde{k}$ to pick up the log divergent piece.  Dropping a linearly divergent term that vanishes by rotational invariance, we find
\be
  &=& \frac{g}{2(1+\gamma)^2} \left[ (- g^2 + 2\alpha^2 \gamma^2) k_1^2 + 2i g \alpha \gamma \frac{\gamma-1}{1+\gamma} \epsilon^{ij} k_2^i k_3^j \right] \int \frac{d^2 \tilde{k}}{(2\pi)^2 }\frac{1}{\tilde{k}^2} .
  \label{eq:vert1}
\ee

Next, there is also a contribution from the bottom two diagrams in Fig. \ref{Fig: YukawaRenormalization}:
\be
&&-2 \gamma \tilde{\alpha}^2 \int \frac{ d \tilde{\omega} d^2 \tilde{k} }{(2\pi)^3} \left[ (-k_1 \cdot \tilde{k})  (-2i \gamma \alpha \epsilon^{ij}  \tilde{k}^i k_3^j -g \tilde{k}^2 ) \left( \frac{1}{ \tilde{\omega}^2 +  \tilde{k}^4 } \right) \left( \frac{ - 1}{i (\omega_3 + \tilde{\omega}) - \gamma(k_3 + \tilde{k})^2} \right) \right. \nn \\
&& \left. \ \ \ \ \ \ \ \ \ \ \ \ \  +k_1 \cdot \tilde{k} (-2i \gamma \alpha \epsilon^{ij} k_2^i  \tilde{k}^j - g \tilde{k}^2) \left( \frac{1}{\tilde{\omega}^2 + \tilde{k}^4 } \right) \left( \frac{ - 1 }{i (\omega_2 + \tilde{\omega} ) - \gamma( k_2 + \tilde{k} )^2 } \right)\right].
\ee
We again close the $\tilde{\omega}$ contour in the upper half-plane to pick up the $\tilde{\omega} = i \tilde{k}^2$ pole.  The subsequent Taylor expansion obtains a log divergent term, which after some simplification can be written as
\be
&&-  \frac{\gamma^2 \tilde{\alpha}^2 }{(1+\gamma)^2} \left[ - 2i  \alpha (1+\gamma) \epsilon^{ij} k_2^i k_3^j +g k_1^2\right]\int \frac{d^2 \tilde{k} }{(2\pi)^2}  \frac{ 1 }{\tilde{k}^2} .
\label{eq:vert2}
\ee
We can combine the corrections  (\ref{eq:vert1}) and  (\ref{eq:vert2}) to read off the necessary counter-terms.  
We find that 
\be
\delta_g &=& -g \frac{ \left( 2 g^2 - 4g (\alpha^2 - \tilde{\alpha}^2)\right)}{4(1+\gamma)^2} \int \frac{d^2 \tilde{k}}{(2\pi)^2 \tilde{k}^2},\\
\delta_{\gamma \alpha} &=& \alpha \gamma \frac{ \left( 2 g^2 (\gamma-1) + 4\tilde{\alpha}^2 \gamma(1+\gamma)^2\right)}{4(1+\gamma)^3} \int \frac{d^2 \tilde{k}}{(2\pi)^2 \tilde{k}^2}.
\ee
and thus the $\beta$ functions are
\be
\beta_{\gamma \alpha} &=& \frac{\partial \gamma \alpha} {\partial \log \mu} = \alpha \gamma^2 \frac{ g^2 + \tilde{\alpha}^2 (1+\gamma)^2}{2\pi(1+\gamma)^3},\\ 
\beta_g &=& \frac{\partial g}{\partial \log \mu} = g \gamma^2 \frac{ \alpha^2 - \tilde{\alpha}^2 }{2\pi(1+\gamma)^2}.
\ee

Note first of all that this explicitly verifies that when $\alpha = \tilde{\alpha}$, we have $\beta_g =0$ and $\beta_{\gamma \alpha} = \alpha \beta_\gamma$, so neither $g$ nor $\alpha$ run, as expected.  A similar, though lengthy, computation involving many diagrams shows that the $\beta$ function for the $\gamma \tilde{\alpha}^2 (\nabla \phi)^2 |\psi|^2$ coupling runs according to
\be
\beta_{\gamma \tilde{\alpha}^2} &=& \gamma^2 \frac{ g^2 \alpha^2 + \tilde{\alpha}^4 (1+\gamma)^2}{2\pi(1+\gamma)^3}.
\ee
This is also in agreement with the non-running of $\alpha$ at the symmetric point $\alpha = \tilde{\alpha}$, since then $\beta_{\gamma \alpha^2} = \alpha^2 \beta_\gamma$.  
  We can combine the above formulae to obtain the $\beta$ function for the coupling $A= 4\gamma(\alpha^2 - \tilde{\alpha}^2)$, which is a measure of how badly the symmetry is broken in the more general case:
 \be
 \beta_A &=& - \frac{A^2}{8\pi(1+\gamma)}.
 \ee
 Thus, the symmetric theory $A=0$ is an attractive fixed point if $A$ is negative in the UV, and a repulsive one $A$ begins from a positive UV value.  This means that our theory with the Cauchy-Riemann symmetry is an attractive IR fixed point for half of the naive UV parameter space.

\section{Adding Spin}
\label{sec:AddingSpin}

We will find it convenient to perform the corresponding calculations using the Wick rotated (i.e. Euclidean) version of the action:
\be
S_E &=& \int dt d^2x 
\left[ \sum_{\sigma=\uparrow, \downarrow}\left( - \psi_\sigma^\dag  \partial_t  \psi_\sigma + \gamma \left| \left(\nabla_i + i\alpha \epsilon_{ij} \nabla^j \phi \right) \psi_\sigma \right|^2 - g \nabla^2 \phi \psi_\sigma^\dag \psi_\sigma \right) 
\right.
\nonumber \\&& + \frac{\lambda}{4} \psi_\uparrow^\dagger \psi_\uparrow \psi^\dagger_\downarrow \psi_\downarrow + \left. \frac{1}{2} \left( \dot \phi^2 +  (\nabla^2 \phi)^2\right)  \right],
\label{eq:ourSpinLagEuc}
\ee
where we have generalized to a two-component spinor $(\psi_\uparrow, \psi_\downarrow)$ and added the quartic $\lambda$ coupling.  We are assuming an $SU(2)$ symmetry connecting $\psi_\uparrow$ and  $\psi_\downarrow$; in the absence of such a symmetry the $\alpha$ and $g$ couplings to the two fermions could be different.   
The primary novelty of spin is the existence of a four-Fermi coupling.  

\subsection{Review of Non-relativistic Spinors in 2+1 dimensions}
\label{NonRelativisticSpinors}

Here we will review relativistic and non-relativistic spinors in 2+1 dimensions, partly to establish our conventions.  
We will choose the following basis for our $\gamma$ matrices:
\begin{eqnarray}
\gamma^1 = \sigma^1, \gamma^2=\sigma^2,  \  \mathrm{ and} \  \gamma^0 = i \sigma^3~.
\end{eqnarray}
Then, it can be easily checked that they satisfy the appropriate anti-commutation relations:
\begin{eqnarray}
\{ \gamma^\mu, \gamma^\nu\} = -2 \eta^{\mu\nu}.  
\end{eqnarray}
The generators of boosts and rotations are given by $S^{\mu\nu} = \frac{i}{4} [ \gamma^\mu, \gamma^\nu]$.  Explicitly, the two boost generators $K^i$ and single rotation generator $J^3$ are 
\begin{eqnarray}
K^1 \equiv S^{10} = \frac{1}{2} \left( \begin{array}{cc} 0 & 1 \\ -1 & 0 \end{array} \right), \qquad
K^2 \equiv S^{02} = \frac{1}{2} \left( \begin{array}{cc} 0 & i \\ i & 0 \end{array} \right), \qquad
J^3 \equiv S^{21} = \frac{1}{2} \left( \begin{array}{cc} 1 & 0 \\ 0 & -1 \end{array} \right), 
\end{eqnarray}
and a general transformation takes the form
\begin{eqnarray}
\Lambda = e^{i \omega_{\mu\nu} S^{\mu\nu}} .
\end{eqnarray}
Let us take the fermion field $\Psi$ to be the following two-component spinor: 
\begin{eqnarray}
\Psi = \left( \begin{array}{c} \psi_\uparrow \\ \psi_\downarrow \end{array} \right).
\end{eqnarray}
  It is conventional and convenient in relativistic theories to define a ``barred'' fermion operator $\bar{\Psi}$ by
\begin{eqnarray}
\bar{\Psi} = \Psi^\dagger i \gamma^0
\end{eqnarray}
since $ U^\dagger i \gamma^0 = i \gamma^0 U^{-1}$.  Then,  in the relativistic case, the only invariant scalar bilinear we can construct (without derivatives) is
\begin{eqnarray}
\bar{\Psi} \Psi &=& - \psi_\uparrow^\dagger \psi_\uparrow + \psi_\downarrow^\dagger \psi_\downarrow .
\end{eqnarray}
Furthermore, the standard quadratic term is
\be
\bar{\Psi}(-p) i \gamma^\mu p_\mu \Psi(p) - m \bar{\Psi} (-p)\Psi(p) = \left( \begin{array}{c} \psi^*_\uparrow (-\omega,-p) \\ \psi^*_\downarrow (-\omega, -p) \end{array} \right)
  \left( \begin{array}{cc} \omega + m & -i (p_x -i p_y) \\ i (p_x + i p_y) & \omega - m \end{array} \right) \left( \begin{array}{c} \psi_\uparrow (\omega, p) \\ \psi_\downarrow(\omega, p) \end{array} \right)
  \label{eq:standardquad}
\ee
Note that at vanishing spatial momentum, the solutions to the dispersion relation are $\omega=m$ with $\psi_\uparrow=0$ and $\omega=-m$ with $\psi_\downarrow=0$, so when $m$ is large compared to $\vec{p}$, the upper component of $\Psi$ is almost all anti-particle.  

In a non-relativistic theory, one integrates out the anti-particle $\psi_\uparrow$, obtaining the following kinetic term for $\psi_\downarrow$:
\be
\LL
  &=& \psi^*_\downarrow (-\omega, -p) \left( \omega - \frac{\vec{p}^2}{2m+\omega} \right) \psi_\downarrow(\omega, p).
  \ee
In the effective theory, $\frac{\vec{p}^2}{2m+\omega}$ is Taylor expanded at large $m$ and subleading terms are treated as interactions.  
Because a relativistic fermion in 2$+$1 dimensions gives us only a spin-up anti-particle and a spin-down particle, the former of which we integrate out, if we want to have both a spin-up and spin-down particle then we must introduce a second relativistic fermion $X=(x_1, x_2)$ with the sign of the mass term reversed.

So far, we have not included any explicit breaking of Lorentz invariance.  
However, in the non-relativistic case, we are free to impose only the rotation part of the Lorentz group.  In 2+1 dimensions, this is just the $U(1)$ group generated by $J_3$, which is why the $\psi_\uparrow, \psi_\downarrow$ basis is natural.  Under this subgroup, we simply have
\begin{eqnarray}
U \psi_\uparrow = e^{ i \omega_{21} /2} \psi_{\uparrow} \qquad U \psi_\downarrow = e^{- i \omega_{21} /2} \psi_{\downarrow} ~.
\end{eqnarray}
Now, we are free to make more bilinears:
\begin{eqnarray}
\psi_\uparrow \psi_\downarrow + \hc , \quad \psi^\dagger_\uparrow \psi_\uparrow, \quad \text{ and  }  \quad \psi^\dagger_\downarrow \psi_\downarrow~.
\end{eqnarray}
This is nothing more than the fact that non-relativistic physics doesn't know about spin-statistics, so the possible mass terms are equivalent to the possible mass terms for two complex scalars $\phi_1$ and $\phi_2$ oppositely charged under a $U(1)$ flavor symmetry: $\phi_1^* \phi_1, \phi_2^* \phi_2, $ and $\phi_1 \phi_2 + c.c.$ 

We can furthermore choose to impose $P$ and/or $T$ if we wish.  $P$ leaves spin invariant, so does not give any interesting constraints on the above bilinears.  $T$ takes $\psi_\uparrow \leftrightarrow \psi^\dagger_\downarrow$, so therefore the mass terms consistent with rotations and $T$ are
\begin{eqnarray}
\psi_\uparrow \psi_\downarrow + \hc , \quad \text{ and } \quad
- \psi_\uparrow^\dagger \psi_\uparrow + \psi_\downarrow^\dagger \psi_\downarrow .
\end{eqnarray}
Finally, it is technically natural to set all fermion bilinear terms to zero except for the 
standard mass term $\bar{\Psi} \Psi$, since in this limit there is an enhanced $SO(2)$ symmetry:
\begin{eqnarray}
\left(\begin{array}{cc} \psi_\uparrow \\ \psi^\dagger_\downarrow \end{array} \right) \rightarrow 
\left(\begin{array}{cc} \psi_\uparrow \cos \theta + \psi^\dagger_\downarrow \sin \theta \\ \psi^\dagger_\downarrow \cos \theta -  \psi_\uparrow \sin \theta\end{array} \right) .
\end{eqnarray}

\subsection{Quartic Vertex Corrections}
\label{app:QuarticVertex}

Let us consider the $\beta$ function for the fermion quartic coupling in the case where we have a 2-component $(\psi_\uparrow, \psi_\downarrow)$ theory.  It turns out that due to power counting, only the $g$ coupling and the $\gamma \alpha^2$ scalar-scalar-fermion-fermion couplings are relevant for the logarithmic divergence of the fermion 4-pt function, so we need only consider diagrams involving these couplings.  Also, note that terms with oppositely directed fermion arrows can give factors of $1/\gamma$, which will be important in the small $\gamma$ limit.

The $g^4$ diagram has a logarithmic divergence and goes as
\be
&& g^4 \INT \frac{\tilde k^8}{(\tilde \omega^2 + \tilde k^4)^2} \left( \frac{1}{(i \tilde \omega - \gamma k^2)^2} + \frac{1}{\tilde \omega^2 + \gamma^2 k^4} \right)
\\
&=&  g^4 \left(  \frac{1+3\gamma + \gamma^2}{2 \gamma (1 + \gamma)^3} \right)
 \int \frac{d^2 \tilde k}{(2 \pi)^2} \frac{1}{\tilde k^2} 
\ee
where the two terms in the first two lines come from the two possible directions for the arrows.
The pure $(\gamma \alpha^2)^2$ diagram has a symmetry factor of $1/2$, but there are $2$ different arrow directions, giving
\be
4\gamma^2 \alpha^4 \INT
\frac{\tilde k^4}{(\tilde \omega^2 + \tilde k^4)^2}
= \gamma^2 \alpha^4  \int \frac{d^2 \tilde k}{(2 \pi)^2} \frac{1}{\tilde k^2}~.
\ee
There are 6 diagrams involving $g$ couplings attached to fermion lines in the $\lambda$ 4-fermion coupling, and we find
\be
&& \lambda g^2 \INT 
\frac{\tilde k^4}{\tilde \omega^2 + \tilde k^4} \left( \frac{4}{(i \tilde \omega - \gamma k^2)^2} + \frac{2}{\tilde \omega^2 + \gamma^2 k^4} \right)
\\
&=& \lambda g^2 \frac{1- \gamma }{\gamma (1 + \gamma)^2} \int \frac{d^2 \tilde k}{(2 \pi)^2} \frac{1}{\tilde k^2}~.
\ee
Finally, there are diagrams involving two $g$ couplings and a $\gamma \alpha^2$, giving
\be
&& 2\gamma \alpha^2 g^2 \INT 
\frac{\tilde k^6}{(\tilde \omega^2 + \tilde k^4)^2} \left( \frac{-1}{i \tilde \omega - \gamma \tilde k^2} +  \frac{-1}{-i \tilde \omega - \gamma \tilde k^2}  \right)
\\
&=& 2\gamma \alpha^2 g^2 \frac{2 + \gamma}{2(1+\gamma)^2} \int \frac{d^2 \tilde k}{(2 \pi)^2} \frac{1}{\tilde k^2} 
\ee
where the two terms come from the two possible directions for the arrows.  
There are also the pure $\lambda^2$ terms
\be
\lambda^2 \INT 
\frac{1}{\tilde \omega^2 + \gamma^2 \tilde k^4} 
= \frac{ \lambda^2}{2 \gamma} \int \frac{d^2 \tilde k}{(2 \pi)^2} \frac{1}{\tilde k^2} ~.
\ee
Thus the total $\beta$ function is
\be
\beta_\lambda 
= \frac{1}{2 \pi} \left( \frac{ \lambda^2}{2 \gamma} + \lambda g^2 \frac{1 }{\gamma (1 + \gamma)^2} + 2\gamma \alpha^2 g^2 \frac{2 + \gamma}{2(1+\gamma)^2}  +  \gamma^2 \alpha^4 + g^4 \left(  \frac{1+3\gamma + \gamma^2}{2 \gamma (1 + \gamma)^3} \right) \right)~.
\ee
Note that in the limit of $\gamma \ll 1$, we have
\be
\frac{1}{2 \pi} \left( \frac{ (\lambda + g^2)^2 }{2 \gamma} - 2 g^2 \lambda\right)~.
\ee
Thus we see that at small $\gamma$, the leading term has a fixed point at $\lambda = -g^2$, although the next term destroys it.

\bibliographystyle{JHEP}
\renewcommand{\refname}{Bibliography}
\addcontentsline{toc}{section}{Bibliography}
\providecommand{\href}[2]{#2}\begingroup\raggedright
\end{document}